\documentclass[%
 reprint,
 amsmath,amssymb,
 aps,
 pra,
]{revtex4-2}

\usepackage{graphicx}
\usepackage{dcolumn}
\usepackage{bm}
\usepackage{amsmath}
\usepackage{physics}
\usepackage{mathtools}
\usepackage{float}
\usepackage[subrefformat=parens]{subcaption}
\usepackage{listings}
\usepackage{here}

\usepackage{xcolor}

\begin{document}

\preprint{APS/123-QED}

\title{Simple loss-tolerant protocol for GHZ-state distribution in a quantum network}

\author{Hikaru Shimizu$^{1}$}
 \email{shimihika2357@keio.jp}
 
\author{Wojciech Roga$^{1}$}
 \email{wojciech.roga@keio.jp}

\author{David Elkouss$^{2,3}$}
 \email{david.elkouss@oist.jp}

\author{Masahiro Takeoka$^{1,4}$}
 \email{takeoka@elec.keio.ac.jp}

\affiliation{%
$^{1}$Department of Electronics and Electrical Engineering, Keio University, 3-14-1 Hiyoshi, Kohoku-ku, Yokohama 223-8522, Japan
} 
\affiliation{%
$^{2}$Networked Quantum Devices Unit, Okinawa Institute of Science and Technology Graduate University, Okinawa,Japan
}
\affiliation{%
$^{3}$QuTech, Delft University of Technology,Lorentzweg 1, 2628 CJ Delft,The Netherlands
}
\affiliation{%
$^{4}$National Institute of Information and Communications Technology (NICT), Koganei, Tokyo 184-8795, Japan
}

\date{\today}

\begin{abstract}
Distributed quantum entanglement plays a crucial role in realizing networks that connect quantum devices. However, sharing entanglement between distant nodes by means of photons is a challenging process primary due to unavoidable losses in the linking channels. In this paper, we propose a simple loss-tolerant protocol for the Greenberger-Horne-Zeilinger state distribution. We analyze the distribution rate under feasible experimental conditions and demonstrate the advantages of rate-loss scaling with respect to direct transmission. Our protocol does not use quantum repeaters and is achievable with current quantum optics technology. 
The result has direct application to tasks such as conference key agreement or distributed sensing. Moreover, it reduces the requirements for implementing distributed quantum error correction codes such as the surface code.
\end{abstract}

\maketitle




Quantum links allow distant nodes to be connected in quantum networks \cite{Wehner2018,Kimble2008,Gisin2007} which are expected to impact such fields as cryptography \cite{Bennett2014,Ekert1991,Bennett1992,Chen2007,Proietti2021,Carrara2023},
high accuracy clock synchronization \cite{Komar2014}, longer-baseline telescopes \cite{Gottesman2012,Khabiboulline2019}, quantum sensing \cite{Eldredge2018, Ge2018, Qian2019, Qian2021}, and quantum computing \cite{Cacciapuoti2020,Rodney2016}. In quantum networks, users can exchange arbitrary quantum states between distant nodes through quantum teleportation \cite{Bennet1993} which consumes shared entanglement. Therefore, bipartite entanglement distribution is one of the most important processes for quantum networks. Multipartite entanglement is often distilled from copies of the bipartite one and became a standard procedure both in experimental and theoretical studies. However, generating multipartite entangled states directly, also shows advantages in specific cases \cite{Epping2017,Yamasaki2018}. 

\begin{figure}[h]
    \centering
    \includegraphics[width=1\linewidth]{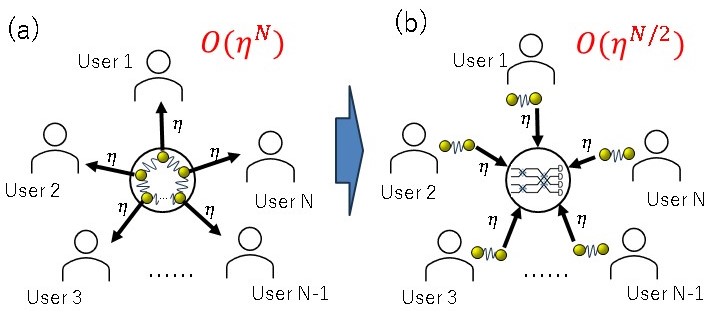}
    \caption{\label{fig1} \raggedright Generating multipartite entangled states in a star network. (a) Direct transmission protocol. (b) Proposed protocol with appropriate interference and conditional measurement in the central node. Black lines indicate optical fibres of equal length. $\eta$ is the transmittance of the links.}
\end{figure}

In typical networks, quantum entanglement is shared by means of transmitted photons. Considering star networks, as in Fig.~\ref{fig1}, the simplest protocol distributing multipartite entanglement consists in generating a desired quantum state locally at the central node and transmitting it by photons to each user node as in Fig.~\ref{fig1}(a). 
This solution was demonstrated experimentally \cite{Proietti2021} and proven feasible, however, its efficiency is strongly limited by the photon loss. Indeed, here, $N$ photons, where $N$ is the number of end-node users, must simultaneously successfully survive transmission through the lossy channels. Therefore the distribution rate scales as $O(\eta^N)$, where $\eta$ is the probability that one photon successfully passes through the channel of a given length \cite{Takeoka2014, Pirandola2017}. 
This protocol is referred to as the 'direct transmission protocol' due to the absence of any repeater-like operations. 
In principle, a perfect quantum repeater with quantum memory at the central node can be implemented in a protocol achieving the distribution rate that scales as $O(\eta)$ \cite{Duan2001, Sangouard2011, Azuma2021}. However, although quantum repeaters technology has made significant progress in recent years and some experimental demonstrations have been performed \cite{Munro2015,Briegel1998,Hasegawa2019,Krutyanskiy2023,Azuma2022}, its implementation in multi-partite scenarios is still beyond the reach of current technology. Indeed, even two-party implementation is still challenging. Therefore, how to achieve an advantageous multi-partite entanglement generation rate with current technology remains a major unsolved issue which we address in this letter.

In \cite{Roga2023} the authors proposed a simple loss-tolerant protocol to distribute the W states and the Dicke states at the rate of $O(\eta)$ and $O(\eta^M)$, 
respectively, with $M<N$. 
It was, however, not known if it could be generalized to generate stronger correlated states like the Greenberger-Horne-Zeilinger (GHZ) states:
\begin{equation}
    \label{GHZ}
    \ket{{\rm GHZ}}_N = \frac{1}{\sqrt{2}}(\ket{000...0}+\ket{111...1}),
\end{equation}
which are of capital importance for tasks such as conference key agreement \cite{Chen2007,Proietti2021,Carrara2023} or quantum sensing \cite{Eldredge2018, Ge2018, Qian2019, Qian2021}. In the present letter, we show that it is, indeed, possible. We establish a new specific protocol, and investigate its performance under realistic experimental conditions. Namely, we calculate the rate and fidelity of GHZ states generation in a star network considering realistic accessible sources and imperfections of measurement devices. We show that our protocol significantly beats the direct transmission protocol. Our protocol has immediate impact on well-known protocols based on the GHZ states. Moreover, we propose its application to improve the implementation of stabilizer codes between distributed qubit memories. 
 
\begin{figure}[h]
    \centering
    \includegraphics[width=1\linewidth]{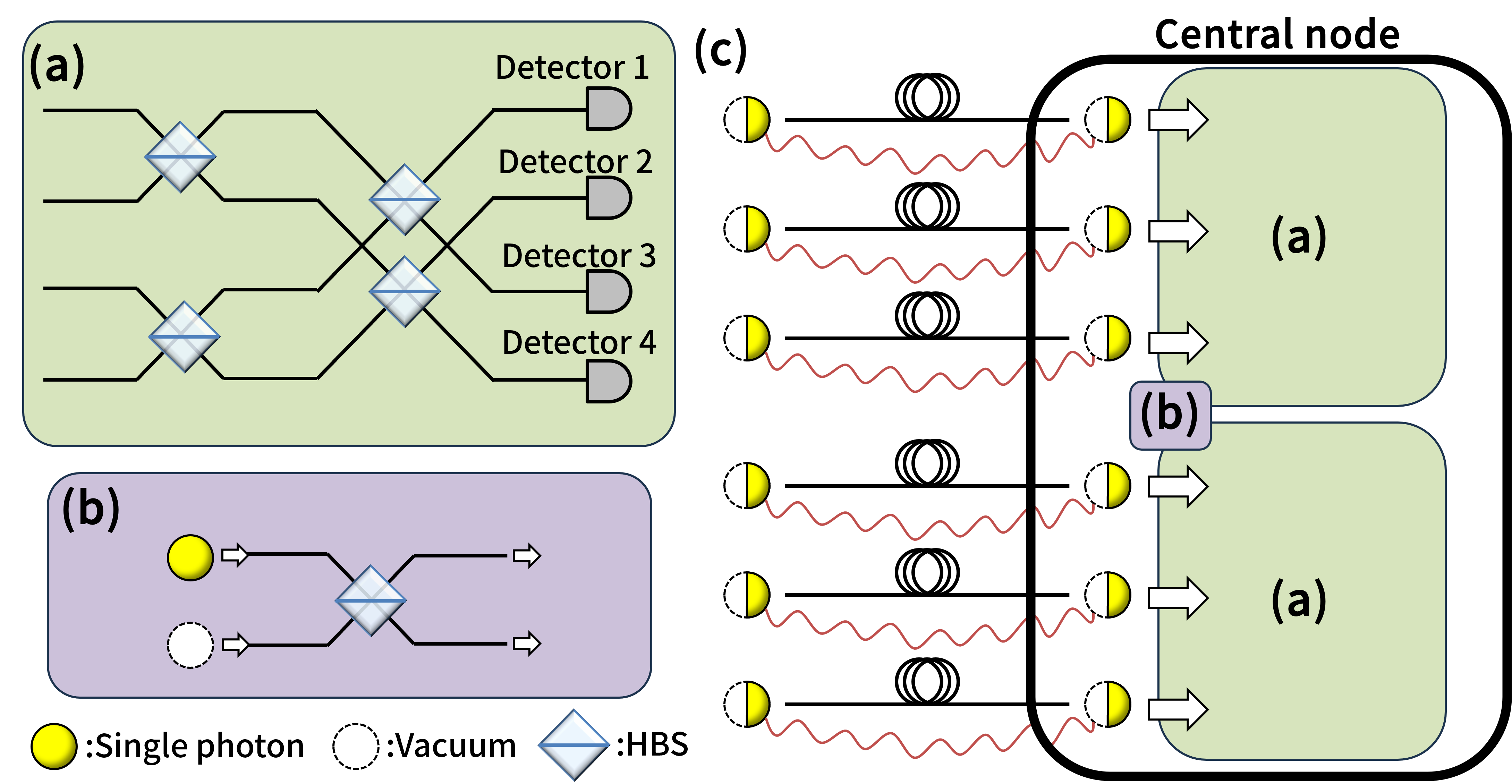}
    \caption{\label{fig3} \raggedright (a) Central node 4-mode circuit generating state $|\rm GHZ\rangle_{4}$ in users' nodes $X_1,...,X_4$ when specific pair of bucket detectors detect photons. (c) Central node circuit generating $|\rm GHZ\rangle_6$ at users' nodes when specific configuration of four bucket detectors detect photons. The central node uses adjacent modes of building blocks (a) and single photon source (b) to connect the blocks. The setup can be extended to generate $|\rm GHZ\rangle_N$ with arbitrary $N$ by adding further building blocks and connect their adjacent modes analogously.} 
\end{figure}

{\it Protocol.} Consider a star network with $N$ users equally distant from the central node. The goal is to generate the GHZ-state among the users. We assume that each user locally generates the state
\begin{equation}
    \label{initial-ideal}
    \ket{\psi}_{X_iX'_i} = a\ket{00}_{X_iX'_i} + b\ket{11}_{X_iX'_i},
\end{equation}
where $|a|^2 + |b|^2 = 1$. Here, $X_i$ and $X'_i$ denote the subsystems retained by the $i$-th user node, which may be photonic or atomic system, and transmitted to the central node -- flying qubit in photon number basis, respectively. 
At the central node, the flying qubits interfere with each other, after which a measurement in the Fock basis is performed. The goal is to find an optical circuit which makes flying qubits interfere appropriately to conditionally transform the retained qubits into the GHZ-state depending on the result of the central node measurement. The role of the central node resembles that of the Bell-state measurement in entanglement swapping. Essentially, we can characterize our objective at the central node as an extension of the multipartite Bell measurement. As this protocol creates the superposition state by erasing information about the origin of the detected photons, the final state must consists of terms with the same total number of photons. So, let us consider the following GHZ state 
\begin{equation}
    \label{GHZ_fock}
    \ket{\rm GHZ}_{4} = \frac{1}{\sqrt{2}}(\ket{1010}_{X_1X_2X_3X_4}+\ket{0101}_{X_1X_2X_3X_4}),
\end{equation}
where the indexes indicate the users. In this letter, we define a GHZ state as every state equivalent to (\ref{GHZ}) under local operations which 
can be identified by the measurement pattern. We will specify particular realizations if necessary. 
State $\ket{\rm GHZ}_{4}$ in subsystems $X_1,...,X_4$ can be approximately realized by inputting states (\ref{initial-ideal}) in the interferometer in Fig.~\ref{fig3}(a) and observing single-photon detections in appropriate pairs of modes. 
Specifically: detections in modes $(1,2)$ or $(3,4)$ generate $|\Phi\rangle=(\pm |1010\rangle\mp|0101\rangle)/\sqrt{2}$;  $(1,3),(2,4)$: $|\Phi\rangle=(\pm |1100\rangle\mp|0011\rangle)/\sqrt{2}$; and $ (1,4), (2,3)$: $|\Phi\rangle=(\pm |1001\rangle\mp|0110\rangle)/\sqrt{2}$, when there are no channel loss and no experimental imperfections. 
Note that the circuit in Fig.~\ref{fig3} was also used in \cite{Roga2023} to generate W- and Dicke-states.

To generate GHZ states with $N>4$, we use circuit from Fig.~\ref{fig3}(a) as a building block. However, to obtain larger GHZ states this circuit is not extended trivially. 
For that, we propose to use setup as in Fig.~\ref{fig3}(c). Here, copies of circuit (a) are connected through the edge modes by means of auxiliary circuits with a single photon sources as in Fig.~\ref{fig3}(b). This method succeeds if one of the following detection patterns $(1,2), (1,3), (1,4), (2,3), (2,4), (3,4)$ is signalized in each of the 4-mode building block circuits. Thus, in the circuit in Fig.~\ref{fig3}(c) we have $36$ four-photon detection patterns that up to known local unitaries lead to $|\rm GHZ\rangle_6$ in six user nodes. 
By analogously connecting more 4-mode circuits with adjacent auxiliary modes the circuit can be expanded to generate $|\rm GHZ\rangle_N$ with arbitrary $N$. The required resource budget consists of: $\frac{N}{2}-1$ four-mode building blocks, $N-4$ auxiliary modes, $2N-4$ detectors, and $\frac{N}{2}-1$ single photon sources.


To consider loss in each channel, we introduce ancillary modes $E_i$ for each subsystem $X'_i$ and describe photon loss as the interaction between mode $X'_i$ and $E_i$ via a beam splitter.  The transmittance of the beam splitter and the channel are assumed to be equal to the link power transmittance $\eta$.
The standard quantum optics algebra implies that the conditional state with the target detection pattern is going to be
\begin{equation}
    \label{actual}
    \rho = \alpha\ketbra{\Phi} + \sum \beta_i\ketbra{\phi_i},
\end{equation}
where $\alpha+\sum \beta_i=1$ and \{$\ket{\phi_i}$\} are undesirable states stemming from the photon losses. 
Since we can set $|\alpha|\gg|\beta|$ by properly choosing $a$ and $b$, the above state approximates (\ref{GHZ_fock}) or its local equivalence \cite{supplementary}.
The probability of getting one of the target detection patterns is as follows:
\begin{equation}
    \label{rate}
    R = 3^{\frac{N}{2}-1}\left(\frac{1}{2}\right)^{N-4}\eta^{\frac{N}{2}}b^{N} [a^2+b^2(1-\eta)]^{\frac{N}{2}},
\end{equation}
while the fidelity of the resulting state in retaining subsystems $X$ with the target GHZ state, $F \equiv \sqrt{\bra{\Phi}\rho_X\ket{\Phi}}$, reads
\begin{equation}
    \label{fidelity}
    F = \sqrt{\frac{a^{N}}{(a^2+b^2(1-\eta))^{\frac{N}{2}}}}.
\end{equation}
In the limit of large distances, the leading term in the expansion of Eq.~(\ref{rate}) 
around small $\eta$ 
gives us the scaling of the generation rate, i.e., $O(\eta^\frac{N}{2})$. This scaling agrees with the intuition that in our protocol only half of detectors must detect a single photon. So, $\frac{N}{2}$ photons must be transmitted successfully which gives the GHZ generation rate scaling. 
Notice that the scaling is advantageous with respect to 
the direct transmission protocol with scaling $O(\eta^{N})$. 
Note that the greater the loss and the more users, the greater the advantage of our protocol over the direct transmission.

From Eq.~(\ref{fidelity}), we find $a$ in the input states (\ref{initial-ideal}) that leads to given fidelity,
\begin{equation}
    \label{fixa}
    \begin{split}
        &a^2 = \frac{(1-\eta)F^{\frac{4}{N}}}{1-\eta F^{\frac{4}{N}}}.
    \end{split}
\end{equation}
Increasing $a$ which is the contribution of vacuum in the source states improves fidelity but reduces the rate. Therefore, there is a trade-off between the two quantities. We observe, however, that our protocol have the advantage of rate-loss scaling even if the fidelity is fixed close to $1$ \cite{supplementary}. 

Let us also clarify the procedure in the case when  
the number of user nodes is odd, i.e., $2k-1$ $(k \in \boldsymbol{N})$. Then, one should use the optical circuit for $2k$ user nodes and input coherent states, approximated as 
 \begin{equation}
     \label{coherent}
     \ket{\textrm{coherent}}=\frac{\ket{0}+\alpha\ket{1}}{1+|\alpha|^2},
 \end{equation}
into the modes which are not used. Here, $\alpha$ must be fixed equal to the probability of a photon entering other modes. Then the distribution rate is estimated as $O(\eta^k)$, which shows the same advantage of the protocol over the direct transmission.


In realistic experimental setups one could use the input states (\ref{initial-ideal}) ideally generated with single photon sources such as trapped ions \cite{Walker2018,Kobel2021}, NV centers \cite{Rodiek2017}, or quantum dots \cite{Senellart2017}. 
Another source approximating (\ref{initial-ideal}) is an optical system with spontaneous parametric down conversion (SPDC), which is feasible with the current technology and even useful for some applications.
It generates the two-mode squeezed vacuum (TMSV) state in the signal and idler modes, $S$ and $I$, 
\begin{equation}
    \label{tmsv}
    \ket{\textrm{TMSV}} = \sqrt{1-\lambda^2}\sum_{n=0}^\infty\lambda^n\ket{n}_{S}\ket{n}_{I}.
\end{equation}
Here, $\ket{n}$ indicates the $n$-photon Fock state, $\lambda=\tanh{r}$, and $r$ is the squeezing parameter. We refer to the squeezing level in dB as $20\textrm{log}_{10}e^r (\textrm{dB})$.
The multi-photon terms in TMSV 
cause reduction of fidelity since we can not distinguish if two photons come from the same or different sources. 
To avoid this, we use the heralded single-photon by measuring photon number of the signal mode and the initial state is prepared by splitting it by an asymmetric beamsplitter \cite{supplementary}. 
We assume detector efficiency of $80\%$ with dark count rate of $10^{-6}$. 
The rate and fidelity under these conditions can be calculated by using Gaussian states formalism \cite{Weedbrook, Braunstein, Ferraro} with a library 'thewalrus' \cite{thewalrus}. In this analysis, one attempt means that all users send their subsystems $X'$ to the central node, and the process of generating initial states is not included into the distribution rate. The detailed analysis of rates vs distance for particular cases of TMSV squeezing and fidelity of the generated state is done in \cite{supplementary}. Here,
in Fig.~\ref{fig13}, we only show the optimized generation rate for different values of fidelity. We fix the squeezing level at 0.43 dB and optimize the other parameters to get the highest rate satisfying the fidelity condition. 
The figure shows the advantage of our protocol even with feasible photon sources and realistic experimental parameters.

\begin{figure}[h]
    \centering
    \includegraphics[width=0.5\textwidth]{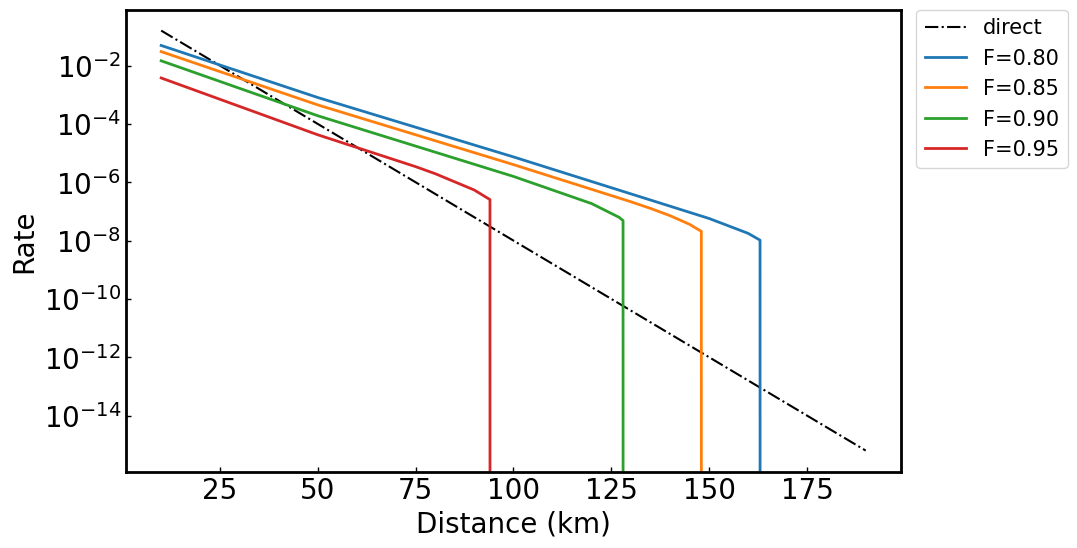}
    \caption{\label{fig13} \raggedright  GHZ generation rate vs distance for different values of the fidelity. Solid lines indicate our protocol and dashed line -- direct transmission. The line with the lowest rate is characterised by the highest fidelity.} 
\end{figure}


{\it Distributed quantum computing architecture.} In~\cite{Nickerson2013}, a distributed architecture for topological quantum computing with noisy network channels was proposed. There, many simple processor cells with a few qubits each
are networked via noisy links. 

\begin{figure}[h]
    \centering
    \includegraphics[width=0.45\textwidth]{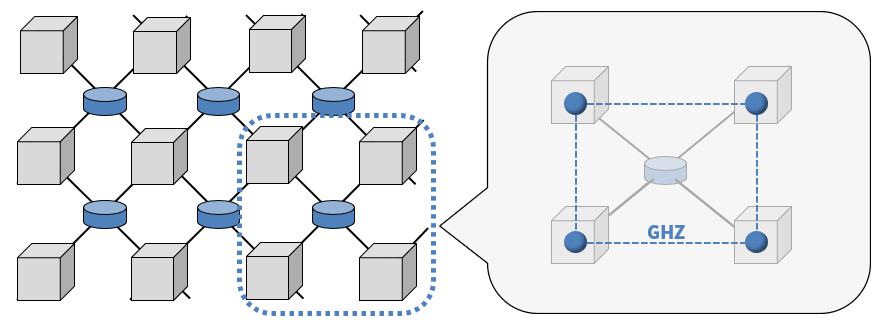}
    \caption{\label{fig14} \raggedright Scheme of distributed 2D surface code architecture. Gray boxes are the cells consisting of small number of qubits. Blue cylinders are the central nodes connected via fibres (black lines) to four nearby cells. The nearest four cells share a GHZ-state, which plays the same roll as the stabilizer in surface code.} 
\end{figure}

The scheme of this architecture is shown in Fig.~\ref{fig14} with a 2D surface code implementation. In this architecture, each data qubit is placed in a different network cell. 
The main difference with respect to the standard surface code architecture is that instead of using adjacent auxiliary qubits for the stabilizer measurements,  4-partite GHZ states are distributed to neighbouring four cells to perform the stabilizer measurements among them. In consequence, the main challenge for such an architecture is the ability to distribute high quality GHZ states at a high rate. 

To address the GHZ distribution problem,  the solution proposed in~\cite{Nickerson2014} has two steps: first, distribute high-quality bipartite states with an efficient photonic linking protocol, called the 'extreme photon loss' (EPL) protocol~\cite{Campbell2008}; and second, fuse and purify the bipartite entanglement to create high-quality GHZ states. With long-enough coherence times~\cite{deBone2024}, this combination has noise thresholds comparable with monolithic architectures even in the presence of link losses and noise.

Essentially, the EPL protocol~\cite{Campbell2008}, is a purification protocol acting on two entangled states generated sequentially. A controlled-NOT (CNOT) operation is performed locally by each party and then one of the states is measured in the $Z$ basis. 
Provided that both measurement outcomes are 1, the remaining state turns out to be an ideal Bell state if no other errors appear. Although, in practice, due to experimental imperfections, it is still a noisy Bell state which should be distilled further. 

\begin{figure}[h]
    \centering
    \includegraphics[width=0.35\textwidth]{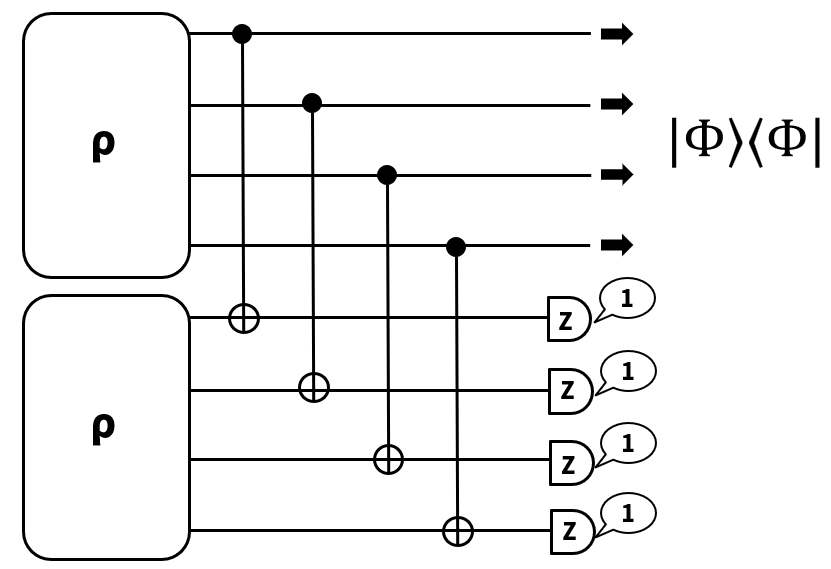}
    \caption{\label{fig16} \raggedright GHZ-state purification of our protocol.} 
\end{figure}

Here, we generalize the bipartite purification in the EPL protocol to more than two end-users (see Fig.~\ref{fig16}) showcasing the feasibility of our protocol to prepare high-quality GHZ states. 
In consequence, our protocol can replace the two-step process of bipartite entanglement generation and GHZ purification. A drawback of fusing bipartite entanglement, is that even in the absence of further purification two-qubit gates are needed, severely impacting performance~\cite{deBone2024}. 
Our protocol allows for a direct generation of the GHZ state among the four neighbouring cells, sidestepping the need of two-qubit gates for fusion. 
Quantitative evaluation of this advantage requires intense numerical simulations and deserves separate research. 

Let us explain how to generalize the EPL protocol. As we mentioned, the appropriate patterns of photodetections in the central node signal the generation of a mixed state (\ref{actual}) in our protocol. 
In particular, if detector 1 and 2 in the central node simultaneously click, the ideal term $\ket{\Phi}$ is $\ket{\Phi} = (\ket{1010}-\ket{0101})/\sqrt{2}$
and the unwanted terms $\ket{\phi_i}\in \{\ket{1011},\ket{1110},\ket{0111},\ket{1101},\ket{1111}\}$. 
We assume that we have two copies of $\rho$ with the same ideal $|\Phi\rangle$ from (\ref{actual}) and perform purification just like the Bell-state purification in the EPL protocol -- 
each cell locally applies CNOT operation to the part of $\rho$ and measures the target qubits in the $Z$ basis. As can be seen form $\ket{\Phi}$ and $\{\ket{\phi}\}$, there is no pair to transform the target qubits into $\ket{1111}$ other than $\{control:\ket{0101},target:\ket{1010}\}$ and vice versa. Thus, like the EPL protocol, we can eliminate the unwanted terms 
by selecting the cases with measurement results '1' in all cells. The probability of getting this result is $P = 0.5\alpha \leq 0.5$. See \cite{supplementary} for details.
The higher $\alpha$, the higher success probability of purification. On the other hand, to get high $\alpha$, the probability that a cell emits a photon must be small which implies longer time to distribute one $\rho$. So, to get the highest distribution rate including the purification process, one needs to optimize over the coefficients of the initial state. 

Concluding, we propose to simplify the original protocol \cite{Nickerson2014} by replacing the process of Bell state generation, purification, and the GHZ state distillation by our direct GHZ generation protocol and distillation by the procedure described above. 




{\it Summary.} In this letter, 
we propose a practical, loss tolerant protocol that directly generates GHZ states in a lossy star network with realistic experimental conditions. The protocol beats the rate of the direct transmission protocol and shifts the common paradigm in which the multipartite entangled states are distilled from bipartite Bell states. 
We analyze our protocol and give the rigorous formulas for the generation rate and fidelity assuming locally produced Bell states as the input to the protocol. We also show how the protocol performs when the input states are TMSV. 
We found that multiple photons in the source, photon loss in channels and dark counts in detectors are the main factors that degrade the fidelity, where the effect of dark counts is particularly significant in longer links. 
Our protocol can directly impact the known GHZ based protocols such as conference key agreement or distributed quantum sensing, which are widely discussed in the literature. Moreover, here we propose its application for distributed codes in quantum computing. We discuss the advantage, leaving a detailed analysis for future work. 
Among future research direction we can indicate the investigation of the effect of phase noise in the transmission process like in \cite{Caspar2020}. 
Moreover, the letter paves the way to design new protocols of distribution of general multipartite entanglement states in quantum networks. 

\begin{acknowledgments}
H.S., D.E. and M.T. are supported by JST Moonshot R\&D Grant No. ~JPMJMS226C. W.R. and M.T. are supported by JST Moonshot R\&D Grant No.~JPMJMS2061. W.R. and M.T. are also supported by JST COI-NEXT Grant No.
JPMJPF2221.
\end{acknowledgments}

\clearpage
\widetext
\begin{center}
   \textbf{\large Supplemental Materials : Simple loss-tolerant protocol for GHZ-state distribution in a quantum network} 
\end{center}
\setcounter{equation}{0}
\setcounter{figure}{0}
\setcounter{table}{0}
\setcounter{page}{1}

\section{The GHZ state distribution protocol}

This section describes our loss-tolerantGreenberger-Horne-Zeilinger (GHZ) state distribution protocol in detail. 
First, we describe the four-partite GHZ state distribution in a lossless star network and then generalize it to the scenario with artibrary number of parties including channel losses.

\subsection{Four-partite GHZ distribution}

\begin{figure}[h]
    \centering
    \includegraphics[width=0.6\textwidth]{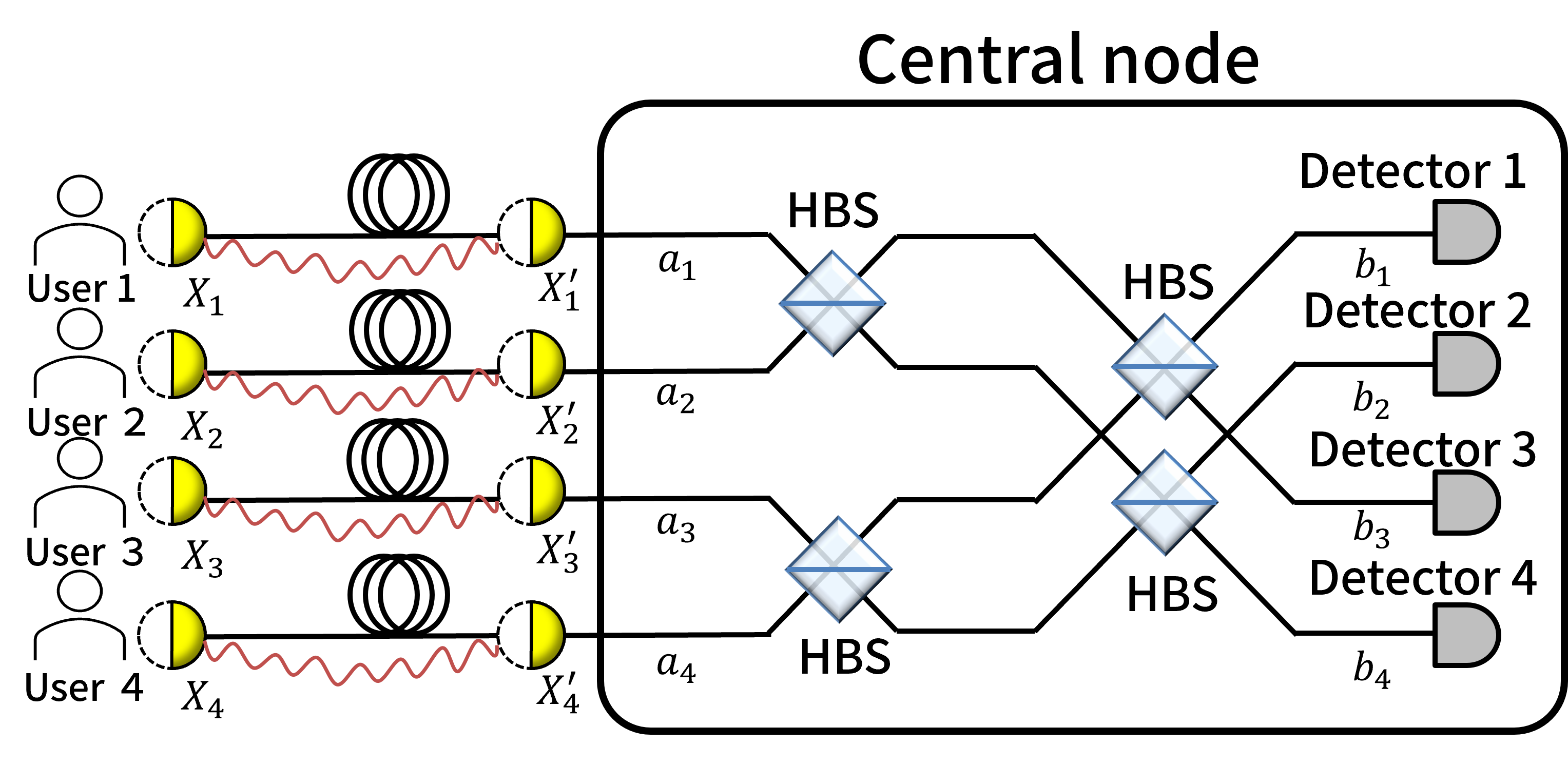}
    \caption{\label{4GHZ_supp} \raggedright Loss tolerant GHZ distribution protocol for four users. HBS: half-beamsplitter.}   
\end{figure}

The $N$-partite GHZ state is defined as a state that up to known local unitary transformations can be written as 
\begin{equation}
    \label{GHZ_s}
    \ket{\rm GHZ}_N = \frac{1}{\sqrt{2}}(\ket{000...0}+\ket{111...1}).
\end{equation}
In this subsection, we discuss how to generate four-mode GHZ state
\begin{equation}
    \label{GHZ_fock_s}
    \ket{\rm GHZ}_{4} = \frac{1}{\sqrt{2}}(\ket{1010}_{X_{1}X_2X_3X_4}+\ket{0101}_{X_{1}X_2X_3X_4}),
\end{equation}
using the circuit of Fig.~2(a) in the main text (see also Fig.~(\ref{4GHZ_supp})).
We assume that each end-node user locally generates Bell state
\begin{equation}
    \label{initial-ideal_s}
    \ket{\psi}_{X_iX'_i} = a\ket{00}_{X_iX'_i} + b\ket{11}_{X_iX'_i},
\end{equation}
where $|a|^2 + |b|^2 = 1$, and sends the subsystem $X'_i$ to the central node. In this subsection, for the sake of clarity of the presentation, we assume no loss. The scenario with losses and other imperfections is discussed in the next subsection.

The initial state of the whole system can be described as the product
\begin{equation}
    \label{initial_s}
    \begin{split}
        \bigotimes_{i=1}^4 \ket{\psi}_{X_iX'_i}&=\bigotimes_{i=1}^4 (a\ket{00}_{X_iX'_i}+b\ket{11}_{X_iX'_i})\\
        &=\ket{\Psi}_{X_{1}X_2X_3X_4X'_{1}X'_2X'_3X_4}.
    \end{split}
\end{equation}
 At the central node, the subsystem $X'_1...X'_4$ undergoes a unitary transformation induced by the four-mode circuit 
 and is measured by a ray of photodetectors. The circuit consists of two rays of half-beamsplitters. Each beamsplitter is described as a unitary transformation of the input photon creation operators
 \begin{equation}
    \label{HBS_s}
    U_\textrm{HBS}
    =
    \frac{1}{\sqrt{2}}
    \begin{bmatrix}
    1&1\\
    1&-1&\\
    \end{bmatrix}.
\end{equation}
If we denote the input and output modes creation operators as $a_i^\dagger$ and $b_i^\dagger$, respectively, the transformation induced by the circuit is
\begin{equation}
    \label{a-b relation_s}
    \begin{bmatrix}
        a_1^\dagger\\
        a_2^\dagger\\
        a_3^\dagger\\
        a_4^\dagger\\
    \end{bmatrix}
    = U_\textrm{4mc}^\dagger
    \begin{bmatrix}
        b_1^\dagger\\
        b_2^\dagger\\
        b_3^\dagger\\
        b_4^\dagger\\
    \end{bmatrix},
\end{equation}
where
\begin{equation}
    \label{uof4mode_s}
    U_\textrm{4mc}
    = U_\textrm{HBS} \otimes U_\textrm{HBS} =
    \frac{1}{2}
    \begin{bmatrix}
    1&1&1&1\\
    1&-1&1&-1\\
    1&1&-1&-1\\
    1&-1&-1&1
    \end{bmatrix}.
\end{equation}
We obtain state (\ref{GHZ_fock_s}) at the user nodes when only two photons arrive at the central node and are detected by appropriate pair of detectors. There are $_4C_2=6$ terms in the shared subsystems $X'_1...X'_4$ of the input state (\ref{initial_s}) that contain two photons. 
These terms are transformed by the four-mode circuit at the central node as follows,
\begin{equation}
    \label{transform_s}
    \begin{split}
        &a_1^{\dagger}a_2^{\dagger}\ket{0000}_\textrm{in}\\
        &\rightarrow{}
        \frac{1}{4}(b_1^{\dagger 2}-b_2^{\dagger 2} +b_3^{\dagger 2}-b_4^{\dagger 2}+2b_1^{\dagger}b_3^{\dagger}-2b_2^{\dagger}b_4^{\dagger})\ket{0000}_\textrm{out}\\
        &a_1^{\dagger}a_3^{\dagger}\ket{0000}_\textrm{in}\\
        &\rightarrow{}
        \frac{1}{4}(b_1^{\dagger 2}+b_2^{\dagger 2} -b_3^{\dagger 2}-b_4^{\dagger 2}+2b_1^{\dagger}b_2^{\dagger}-2b_3^{\dagger}b_4^{\dagger})\ket{0000}_\textrm{out}\\
        &a_1^{\dagger}a_4^{\dagger}\ket{0000}_\textrm{in}\\
        &\rightarrow{}
        \frac{1}{4}(b_1^{\dagger 2}-b_2^{\dagger 2} -b_3^{\dagger 2}+b_4^{\dagger 2}+2b_1^{\dagger}b_4^{\dagger}-2b_2^{\dagger}b_3^{\dagger})\ket{0000}_\textrm{out}\\
        &a_2^{\dagger}a_3^{\dagger}\ket{0000}_\textrm{in}\\
        &\rightarrow{}
        \frac{1}{4}(b_1^{\dagger 2}-b_2^{\dagger 2} -b_3^{\dagger 2}+b_4^{\dagger 2}-2b_1^{\dagger}b_4^{\dagger}+2b_2^{\dagger}b_3^{\dagger})\ket{0000}_\textrm{out}\\
        &a_2^{\dagger}a_4^{\dagger}\ket{0000}_\textrm{in}\\
        &\rightarrow{}
        \frac{1}{4}(b_1^{\dagger 2}+b_2^{\dagger 2} -b_3^{\dagger 2}-b_4^{\dagger 2}-2b_1^{\dagger}b_2^{\dagger}+2b_3^{\dagger}b_4^{\dagger})\ket{0000}_\textrm{out}\\
        &a_3^{\dagger}a_4^{\dagger}\ket{0000}_\textrm{in}\\
        &\rightarrow{}
        \frac{1}{4}(b_1^{\dagger 2}-b_2^{\dagger 2} +b_3^{\dagger 2}-b_4^{\dagger 2}-2b_1^{\dagger}b_3^{\dagger}+2b_2^{\dagger}b_4^{\dagger})\ket{0000}_\textrm{out}.\\
    \end{split}
\end{equation}
Here, we ignore the case in which two or more photons come from one user or two or more photons are detected at one detector. Then, using Eq.~(\ref{transform_s}), we derive states $\ket{\Phi}_{X_1...X_4}$ of the remaining subsystems depending on the detection pattern. Suppose detectors 1 and 2 detect single-photons. After the transmission and measurement, the state in subsystem $X_1...X_4$ becomes
\begin{equation}
    \label{final_no-loss_s}
    \begin{split}
        \ketbra{\Phi}_X &=
        \Trace_{X'} [\bra{1100}_{X'}( \ketbra{\Psi'}_{XX'})\ket{1100}_{X'}]\\
        &= \frac{1}{2}(\ket{1010}-\ket{0101})(\bra{1010}-\bra{0101}),
    \end{split}
\end{equation}
where we use a simplified notation $X_1...X_4\rightarrow X$ and $X'_1...X_4\rightarrow X'$. 

The other detection patterns generate other type of GHZ states. 
That is, detecting one photons at any of two different detectors, one can distribute a GHZ state. 
Table~\ref{table:transform} lists the detection pattern (the detectors clicked) and the resulting states at the end parties. 
\begingroup
\renewcommand{\arraystretch}{2}
\begin{table}[h]
  \caption{States in subsystem $X_1...X_4$ depending on detection pattern in subsystem $X'_1...X'_4$.}
  \label{table:transform}
  \centering
  \begin{tabular}{cc}
    \hline
    Detection pattern & State $\ket{\Phi}_{X_1...X_4}$  \\
    \hline \hline
    1,2 & $\frac{1}{\sqrt{2}} (\ket{1010}-\ket{0101})$ \\
    \hline
    1,3 & $\frac{1}{\sqrt{2}} (\ket{1100}-\ket{0011})$ \\
    \hline
    1,4 & $\frac{1}{\sqrt{2}} (\ket{1001}-\ket{0110})$ \\
    \hline
    2,3 & $\frac{1}{\sqrt{2}} (-\ket{1001}+\ket{0110})$ \\
    \hline
    2,4 & $\frac{1}{\sqrt{2}} (-\ket{1100}+\ket{0011})$ \\
    \hline
    3,4 & $\frac{1}{\sqrt{2}} (-\ket{1010}+\ket{0101})$ \\
    \hline
  \end{tabular}
\end{table}
\endgroup


\subsection{Rate-loss analysis for $N$-partite GHZ distribution in a lossy star network}

If we assume that the power transmittance or the probability that one photon survives one link is $\eta$, the generation rate of $\ket{\Phi}_{X_1...X_4}$ scales as $O(\eta^2)$ because only two photons are needed to survive the links and be detected at the central node.
Note that the loss affects not only the generation rate, but also the fidelity between the final state and the desired GHZ state. Indeed, the final state of the end-node subsystem $X_1...X_4$ is 
a mixed state described as follows
\begin{equation}
    \label{actual_s}
    \rho = \alpha\ketbra{\Phi} + \sum \beta_i\ketbra{\phi_i},
\end{equation}
where $\alpha+\sum \beta_i=1$ and \{$\ket{\phi_i}$\} are states we do not require which appear due to imperfections such as loss, detector dark counts, quantum efficiency, or multi-photon detection. The rigorous derivation of the formulas for the GHZ generation rate and fidelity in terms of the problem variables for arbitrary number of users is presented in the following section. 

In this subsection, we generalize the scenario of the previous subsection by considering an $N$-partite star network taking into account loss of each channel. 
Transmission through optical fibres is modelled a beamsplitter that mixes subsystem $X'_i$ and uncontrollable environment $E_i$ as,
\begin{equation}
    \label{loss-u_s}
    \ket{1}_{X'_i}\ket{0}_{E_i} \rightarrow \sqrt{\eta}\ket{1}_{X'_i}\ket{0}_{E_i} + \sqrt{1-\eta}\ket{0}_{X'_i}\ket{1}_{E_i}.
\end{equation}
Denote this beamsplitting unitary as $U_{loss}$.
Since $E_i$ is uncontrollable, we describe the evolution of subsystem $X'_i$ by averaging over the degrees of freedom of $E_i$ as follows
\begin{equation}
    \label{@-transmission_s}
    \Trace_{E_i}{(U_{\textrm{loss}}\ketbra{\psi}_{X'_i}\otimes\ketbra{0}_{E_i} U_{\textrm{loss}}^\dagger)}.
\end{equation}

\begin{figure}[h]
    \centering
    \includegraphics[width=0.6\textwidth]{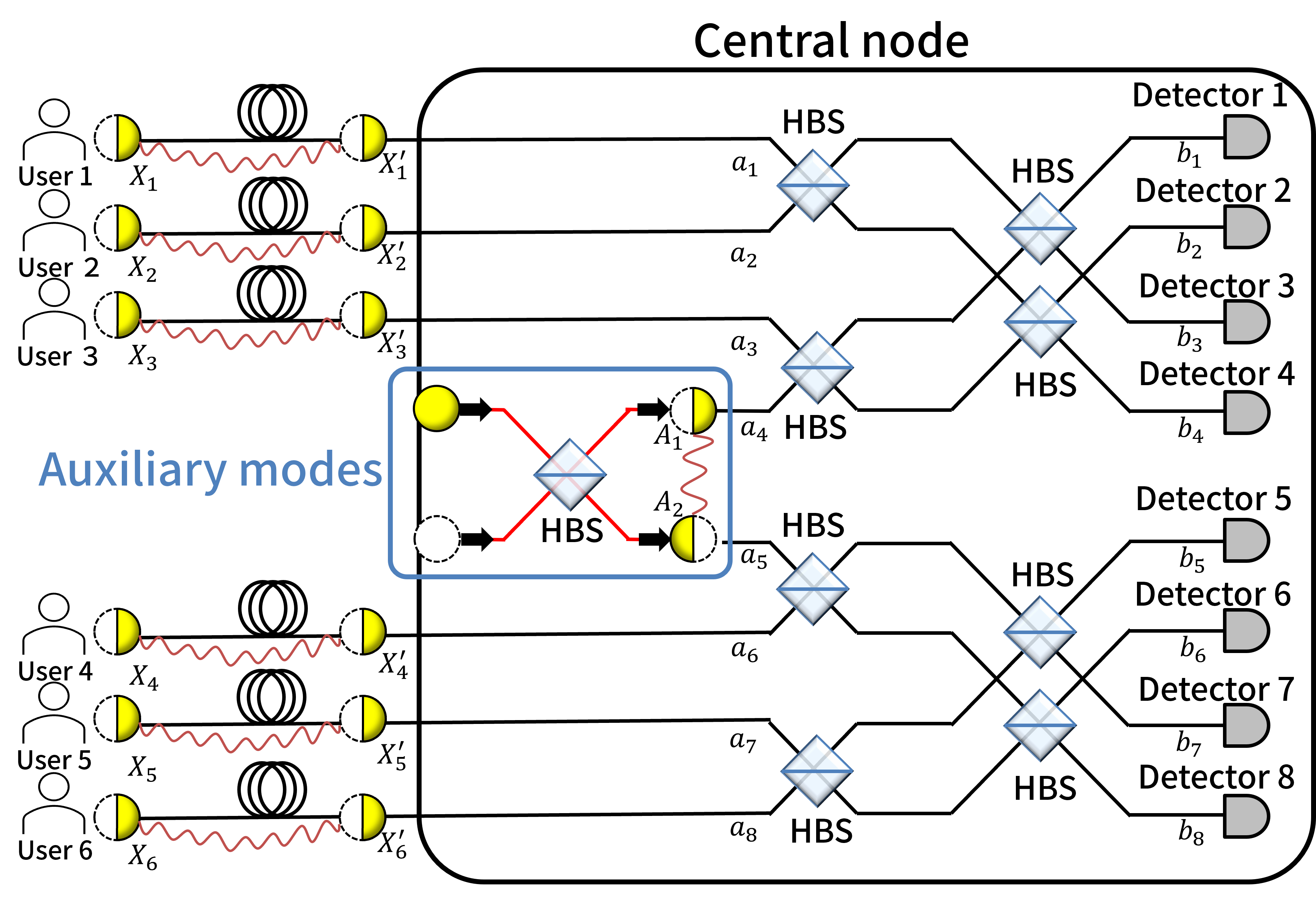}
    \caption{\label{6GHZ_supp} \raggedright Loss tolerant GHZ distribution protocol for six users. HBS: half-beamsplitter.}   
\end{figure}

\begin{figure}[h]
    \centering
    \includegraphics[width=0.5\textwidth]{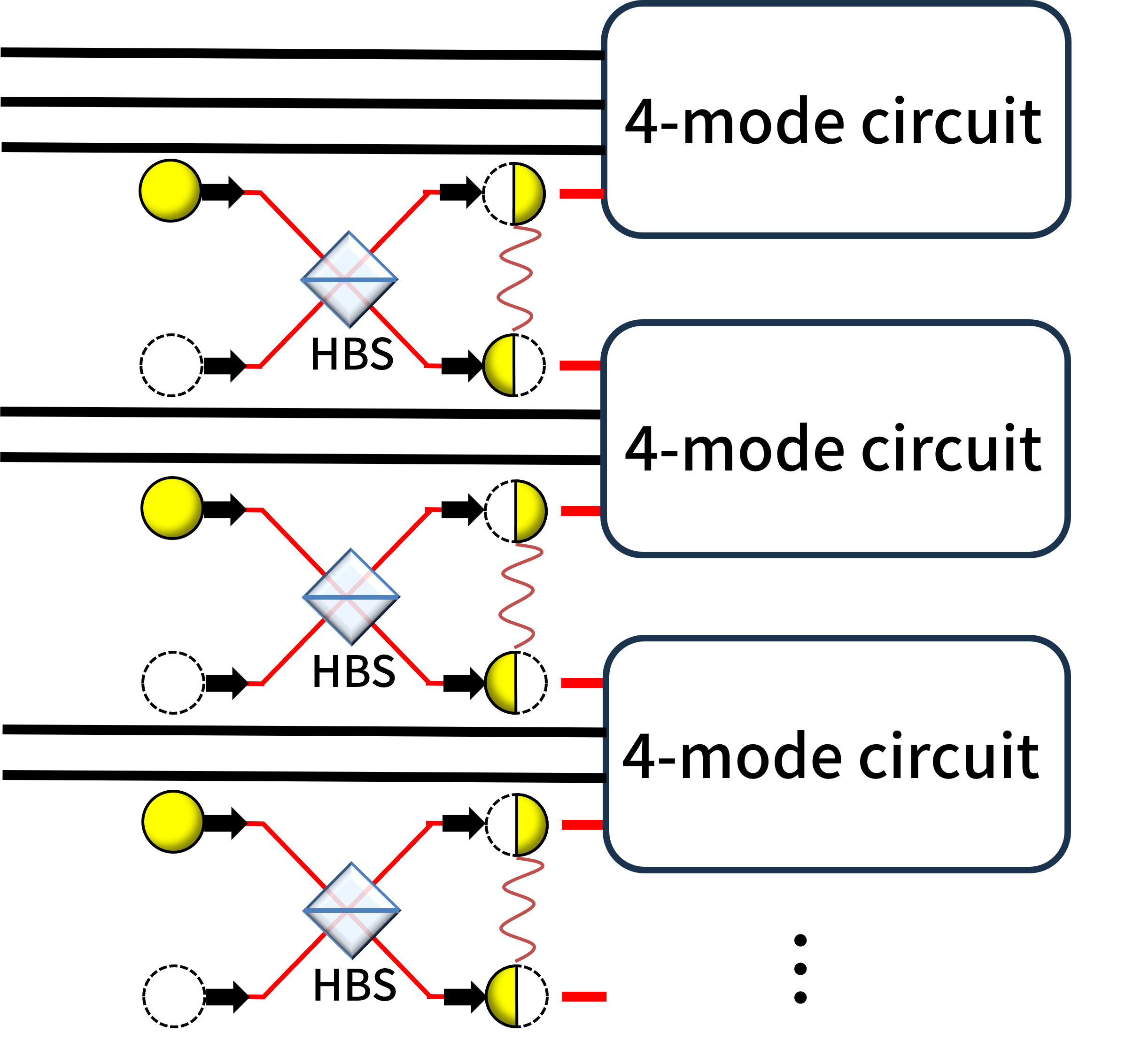}
    \caption{\label{N_GHZ_supp} \raggedright Loss tolerant GHZ distribution protocol for $N$ users. HBS: half-beamsplitter.}   
\end{figure}

Now we consider the case $N>4$ where $N$ is the number of the end users and each user and the central node are connected by a lossy channel with transmittance $\eta$. For simplicity, here we assume $N$ is even. 
Figure 2(c) of the main text shows the setup for $N=6$ as an example. 
The same figure is restated in Fig.~\ref{6GHZ_supp} with more details, where two copies of the circuit of Fig.~\ref{4GHZ_supp} are connected by an auxiliary state having the form of $(|10\rangle+|01\rangle)/\sqrt{2}$. 
It is straightforwardly extended to the setup for $N>6$ by using the circuit of Fig.~\ref{4GHZ_supp} as a building block as illustrated in Fig.~\ref{N_GHZ_supp}. 
Note that for $N$ users, the number of modes at central node is $2N-4$.

Success of the GHZ state distribution is guaranteed by the detection pattern that half of the detectors detect single-photons. 
For any even $N$, due to the symmetry of channel loss and the circuit, the success probability is always the same for all possible successful detection patterns. 
Therefore, without loss of generality, hereafter we consider a particular detection pattern that the first and second detectors in each four-mode building block detect single-photons.  
This means that the state at the detectors are projected onto $\ket{1100...1100}_{X'A}$. 
Consider the initial state, before the channel transmission but including the auxiliary modes, represented by a superposition in the photon number basis. 
The terms that have finite probability to be projected onto $\ket{1100...1100}_{X'A}$ are in the form of $\ket{\psi}_X\ket{1i_11i_21i_3...1i_{N-2}}_{X'A}$ and $\ket{\psi}_X\ket{j_11j_21j_31...j_{N-2}1}_{X'A}$, where $i_l, j_l = 0,1$ $\forall l$. $|\psi\rangle$ are some states depending on $\{i_l\}_l$ and $\{j_l\}_l$. 
In the superposition form, probability amplitudes of these terms are
\begin{equation}
    \label{coef_s}
    Y_m = a^{\frac{N}{2}-m}b^{\frac{N}{2}+m} \times \left(\frac{1}{\sqrt{2}}\right)^{\frac{N}{2}-2},
\end{equation}
where $m=0,1,2,...,N/2$ is the number of modes in which $i_l$ or $j_l$ are 1. 

After the successive detection event, i.e. transmission and projection onto $\ket{1100...1100}_{X'A}$, the initial state turns out to be 
\begin{equation}
    \label{cond_state_s}
    \rho_X = \frac{1}{\sqrt{\mathcal{N}}} \left(
    p_0 Y_0^2 \ketbra{\Phi} + \sum_{m=1}^{\frac{N}{2}} p_m Y_m^2 \sum_k^{2_{N/2}C_m} \ketbra{\phi_{mk}}
    \right), 
\end{equation}
where $|\Phi\rangle$ is the desired GHZ state, $|\phi_{mk}\rangle$ are some states that are orthogonal to $|\Phi\rangle$, and $\mathcal{N}$ is a normalization factor. 
Physically, $m$ is the number of photons that are lost during the transmission and $|\phi_{mk}\rangle$ is an unwanted state that lost $m$ photons during the channel transmission but was projected onto $\ket{1100...1100}_{X'A}$. 
For each $m$, there are $2\times _{N/2}C_{m}$ different $|\phi_{mk}\rangle$, where 2 comes from two types of the terms, $\ket{1i_11i_2...}_{X'A}$ and $\ket{j_11j_21...}_{X'A}$, and each of them has $_{N/2}C_m$ combinations of $\{i_l\}_l$ and $\{j_l\}_l$.
In the main text, $\rho_X$ in Eq.~(\ref{cond_state_s}) is simply expressed as 
\begin{equation}
    \alpha \ketbra{\Phi} + \sum_i \beta_i \ketbra{\phi_i}. 
\end{equation}

The probability that $|\Phi\rangle$ and $|\phi_{mk}\rangle$ are projected onto $\ket{1100...1100}_{X'A}$ are given by 
\begin{equation}
    \label{p-trans_s}
    p_m = \eta^{\frac{N}{2}}(1-\eta)^{m}\times \left(\frac{1}{4}\right)^{\frac{N}{2}-1},
\end{equation}
and combining it with Eq.~(\ref{coef_s}), we have 
\begin{equation}
    \label{coef-fin_s}
    \begin{split}
        &p_m Y_m^2 \\
        &= \eta^{\frac{N}{2}}\left(1-\eta\right)^{m}\times \left(\frac{1}{4}\right)^{\frac{N}{2}-1} \times a^{N-2m}b^{N+2m}\times \left(\frac{1}{2}\right)^{\frac{N}{2}-2}\\
        &= \eta^{\frac{N}{2}}(1-\eta)^{m}\left(\frac{1}{2}\right)^{\frac{3}{2}N-4}a^{N-2m}b^{N+2m}.
    \end{split}
\end{equation}
Then, we have the probability of obtaining the target detection pattern as 
\begin{equation}
    \label{P_s}
    \begin{split}
        P &= 2\sum_{m=0}^{\frac{N}{2}} p_m Y_m^2 \times {}_{\frac{N}{2}}C_m\\
        &= 2\sum_{m=0}^{\frac{N}{2}} \eta^{\frac{N}{2}}\left(1-\eta\right)^{m}\left(\frac{1}{2}\right)^{\frac{3}{2}N-4}a^{N-2m}b^{N+2m} {}_\frac{N}{2}C_m\\
        &= \left(\frac{1}{2}\right)^{\frac{3}{2}N-5}\eta^{\frac{N}{2}}b^{N}\sum_{m=0}^{\frac{N}{2}}{}_\frac{N}{2}C_m\left(a^2\right)^{\frac{N}{2}-m}\left(b^2\left(1-\eta\right)\right)^{m}\\
        &= \left(\frac{1}{2}\right)^{\frac{3}{2}N-5}\eta^{\frac{N}{2}}b^{N} \left[a^2+b^2\left(1-\eta\right)\right]^{\frac{N}{2}}.
    \end{split}
\end{equation}
The total distribution rate $R$ is obtained by multiplying $P$ by the number of combinations of patterns which lead to the GHZ states, i.e.,
\begin{equation}
    \label{rate_s}
    R = 6^{\frac{N}{2}-1}\times P = 3^{\frac{N}{2}-1}\left(\frac{1}{2}\right)^{N-4}\eta^{\frac{N}{2}}b^{N} \left[a^2+b^2\left(1-\eta\right)\right]^{\frac{N}{2}},
\end{equation}
where note that 6 is the number of the detection patterns at each 4-mode building block and $N/2-1$ is the number of the building blocks. 

The fidelity between $\rho_X$ and the ideal GHZ state is also derived as 
\begin{equation}
    \label{fidelity_s}
    \begin{split}
            F & = \sqrt{\bra{\Phi}\rho_X\ket{\Phi}} \\
            & = \sqrt{\frac{4Y_{0}^2p_{0}}{2P}}
            = \sqrt{\frac{a^{N}}{(a^2+b^2(1-\eta))^{\frac{N}{2}}}}.
    \end{split}
    \end{equation}

\begin{figure}[h]
    \centering
    \includegraphics[width=0.55\textwidth]{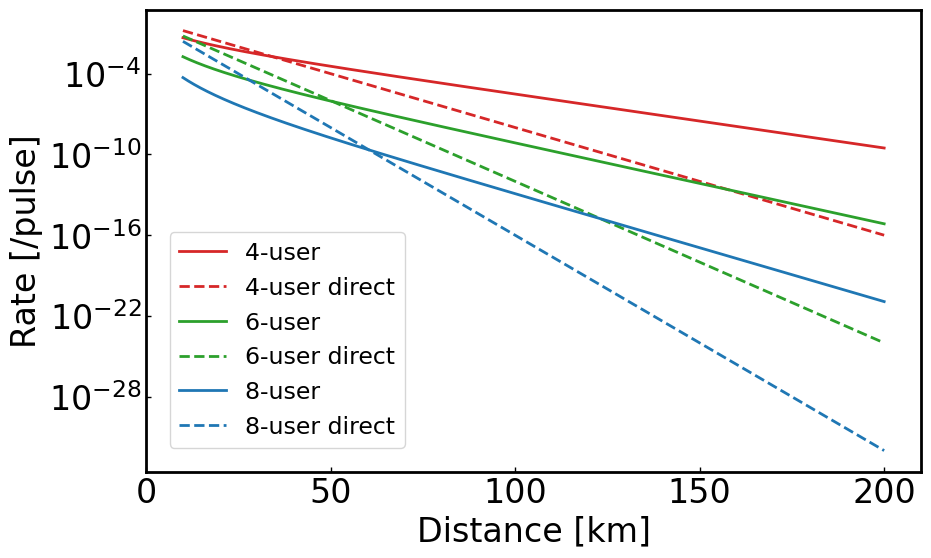}
    \caption{\label{fig7} \raggedright GHZ generation rate vs distance for different numbers of users. The fidelity is fixed at 0.9. The top-red, middle-blue, and bottom-green lines, respectively, represent cases of four, six, and eight users. The solid lines indicate our protocol, while the dashed lines -- the direct transmission, the rate of which is $\eta^{N}$. The lines characterising the scenario with more users show lower rate.}   
\end{figure}

Figure~\ref{fig7} shows the distribution rates $R$ from Eq.~(\ref{rate_s}) of this protocol and the direct transmission protocol for $N=4,6, 8$ user nodes. Here, we assume that the rate of the direct transmission is 
\begin{equation}
    \label{direct_s}
    R_\textrm{direct} \equiv \eta^N.
\end{equation}
\begin{figure}[h]
    \centering
    \includegraphics[width=0.55\textwidth]{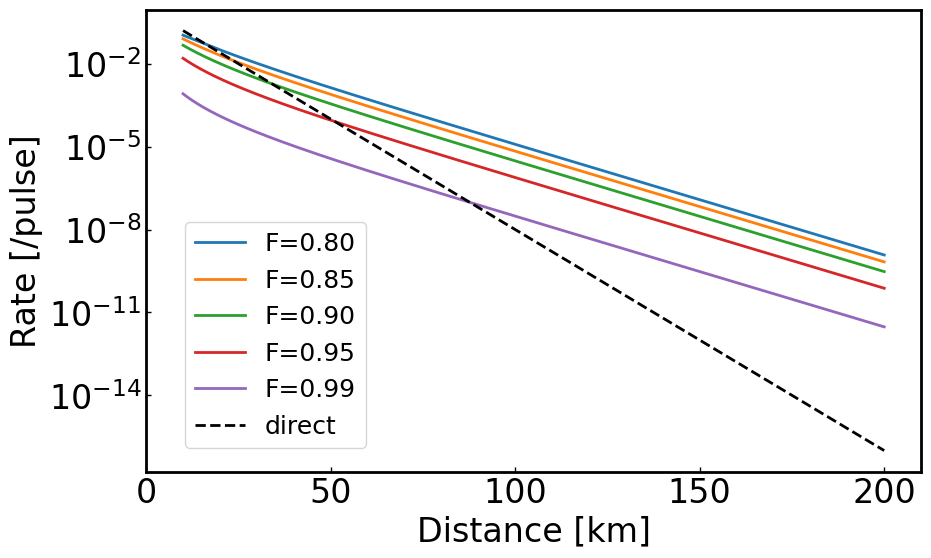}
    \caption{\label{fig8} \raggedright GHZ generation rate vs distance for different fidelities. The fidelity varies between 0.8 and 0.99. The number of user nodes is four. The solid lines indicate our protocol, while the dashed line -- direct transmission. The lines with lower rate are characterised by the higher fidelity.} 
\end{figure}
Figure \ref{fig8} shows the dependence of the rate on the distance for fixed fidelities and comparison with the performance of the direct transmission protocol.
The figures confirm the advantage of the proposed protocol with respect to the direct transmission method especially in the limit of large distances with larger loss and when the number of users increases.

\section{Implementation with spontaneous parametric down conversion sources}
\subsection{Scheme}

\begin{figure}[h]
    \centering
    \includegraphics[width=0.6\textwidth]{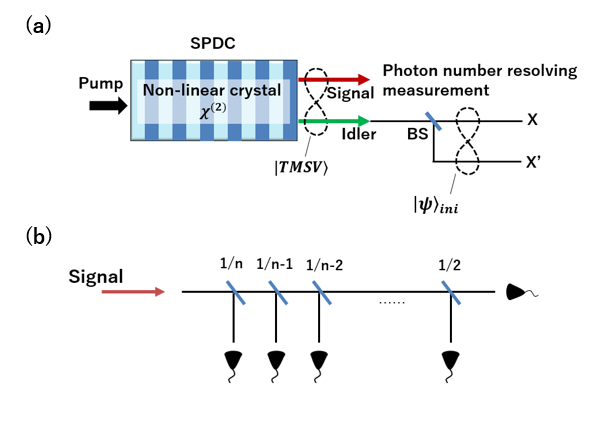}
    \caption{\label{fig9} \raggedright (a) Schematic image of the SPDC source. The signal mode is input to the photon number resolving system (PNRD). The idler mode is split by a beamsplitter (BS) into mode $X$ and $X'$. (b) An example of the setup performing quasi PNRD with threshold detectors. The numbers above each beamsplitter indicate respective reflectance. The click in only one detector indicates a quasi single photon state in the idler mode which is the input in our protocol.} 
\end{figure}

In this section we analyse the performance of our protocol with the initial end-user states generated by spontaneous parametric down conversion (SPDC) source which is widely used in quantum optics experiment due to its availability and efficiency. 
The nonlinear crystal of SPDC generates two-mode squeezed vacuum (TMSV) states in the signal and idler modes denoted by $S$ and $I$ respectively,

\begin{equation}
    \label{tmsv_s}
    \ket{\textrm{TMSV}} = \sqrt{1-\lambda^2}\sum_{n=0}^\infty\lambda^n\ket{n}_\textrm{S}\ket{n}_\textrm{I}.
\end{equation}
Here, $|n\rangle$ is the photon number state, $\lambda=\tanh{r}$, and $r$ is the squeezing parameter. 
As shown in Fig.~\ref{fig9}(a), each end user uses the SPDC source and heralding a single-photon by measuring one of the two modes with a photon-number resolving detector (PNRD). 

The initial state of the protocol is prepared by splitting the heralded single photon, which is expressed as 
\begin{equation}
    \label{ini-herald_s}
    \ket{\psi}_\textrm{ini} = \sqrt{t}\ket{10}_{X_iX'_i} + \sqrt{1-t}\ket{01}_{X_iX'_i},
\end{equation}
where $t$ is the transmittance of the beamsplitter (BS) in Fig.~\ref{fig9}(a). 

While technology of PNRDs are growing, one can also realize a quasi-PNRD by multiplexing threshold detectors, which discriminates only zero or non-zero photons. An example of that is illustrated in Fig.~\ref{fig9}(b).
It allows us to decrease the probability of wrongly counting the photon number. Indeed, in case of the two photon term, the probability that one detector detects two photons in the signal mode is $1/n$, where $n$ is the number of detectors. More generally, if $k$ photons are input to this system, the probability that we mistakenly recognize the state as the single photon state is $(1/n)^{k-1}$. By increasing $n$, one can identify the single photon state in the idler mode with arbitrary precision.

\subsection{Rate and fidelity}

To simulate the results in the realistic experimental conditions we generate the plots of the rate and fidelity of GHZ states distribution with TMSV source described above, with loss and realistic detectors. We assume the loss as in a typical optical fibre which is 0.2~dB/km. The loss rate $\eta$ increases with the distance $L$ (km) as follows
\begin{equation}
    \label{eta_s}
    \eta = 10^{-0.2L/10} = 10^{-0.02L}.
\end{equation}
Detector efficiency is assumed to be $80\%$ with the darkcount probability of $10^{-6}$. 
TMSV is in an infinite dimensional system and is a class of Gaussian state \cite{Weedbrook_s, Braunstein_s, Ferraro_s}.
The numerical analysis is delivered by the Gaussian formalism using the Python library 'thewalrus' \cite{thewalrus_s}.

\begin{figure}[h]
    \centering
    \includegraphics[width=0.5\textwidth]{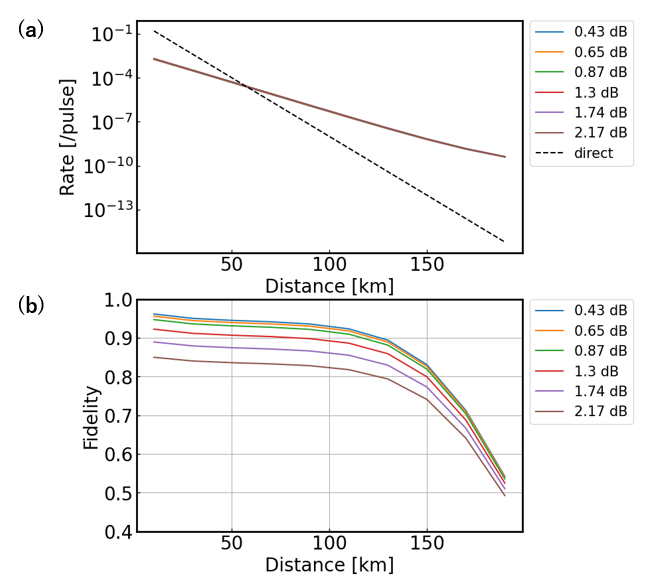}
    \caption{\label{fig10} \raggedright (a) GHZ generation rate vs distance for different squeezing levels. The solid lines indicate our protocol, while the dashed line -- direct transmission.  (b) Fidelity vs distance for different squeezing levels. The lines with smaller fidelity are characterised by larger squeezing.} 
\end{figure}

Figure~\ref{fig10} shows the GHZ states generation rates and the fidelities of the states for different squeezing levels in the four-users case. We fix $t$ in~(\ref{ini-herald_s}) at 0.95. Figure \ref{fig10} (a) suggests that the distribution rate is independent of the squeezing in TMSV. However this reflects the fact that we do not include the input state generation process into the rate. On the other hand, the fidelity strongly depends on the squeezing level. This means that the multi-photon residues in the initial states is an important factor of the fidelity reduction.

\begin{figure}[h]
    \centering
    \includegraphics[width=0.5\textwidth]{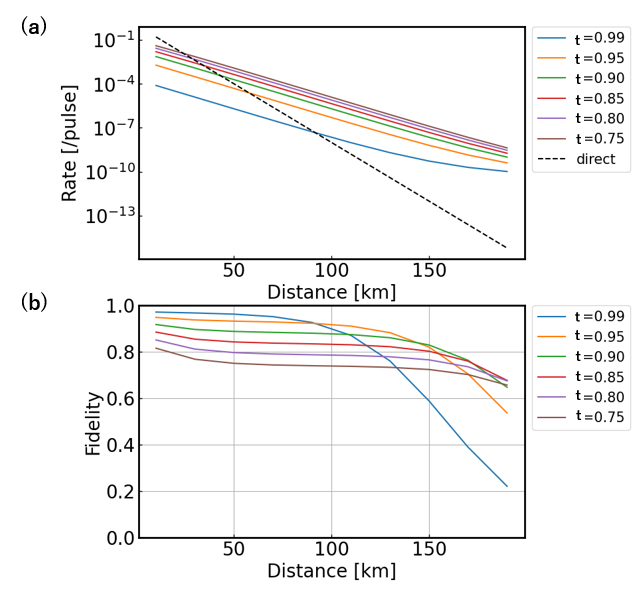}
    \caption{\label{fig11} \raggedright (a) GHZ generation rate vs distance for different values of $t$ in (\ref{ini-herald_s}). The lower rate lines are characterised by larger $t$.  (b) Fidelity vs distance for different values of $t$. The solid lines indicate our protocol, while the dashed line indicates the direct transmission. The lines with higher fidelity for short distances are characterised by larger $t$.} 
\end{figure}

Figure~\ref{fig11} shows the generation rates and fidelities for different values of transmittance $t$ in~(\ref{ini-herald_s}). This figure shows the trade-off between the rate and fidelity. We observe that at some distance the dark counts dominate the detection. Then the generation rate tends to a constant value, however the fidelity drops down abruptly.

\begin{figure}[h]
    \centering
    \includegraphics[width=0.6\textwidth]{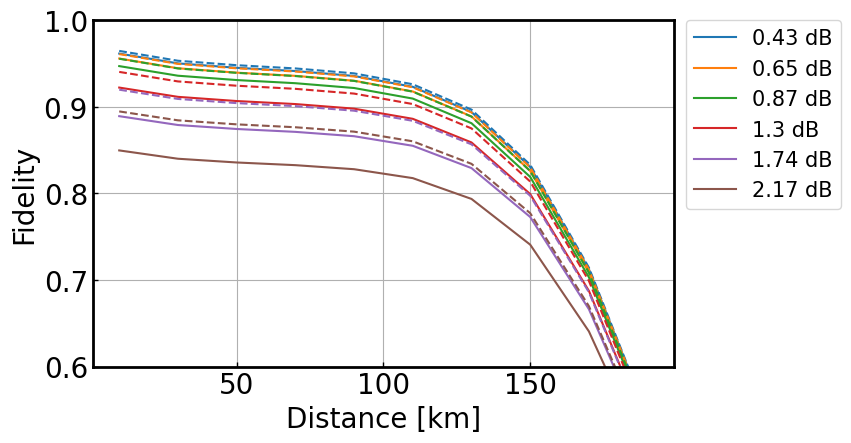}
    \caption{\label{fig12} \raggedright Fidelity vs distance for different levels of squeezing. The solid lines are generated using quasi PNRD with three detectors, while the dashed lines indicate perfect PNRD. The lines showing lower fidelities are characterised by higher squeezing. Here, $t=0.95$.  } 
\end{figure}

To see the effect of PNRD accuracy on the fidelity, we compare the PNRD and the quasi-PNRD with three threshold detectors. 
The result is in Fig.~\ref{fig12}, where we fix $t$ at 0.95. 
The dashed lines indicate the PNRD and the solid lines are for the quasi-PNRD with three detectors. 
We observe that the fidelities are almost the same for these two cases, especially in the small squeezing regime. 
This is because the smaller squeezing, the smaller probability of getting two or more photon terms in the signal and idler modes.

\newpage
\section{Purification of GHZ for the distributed surface codes}

\begin{figure}[h]
    \centering
    \includegraphics[width=0.7\textwidth]{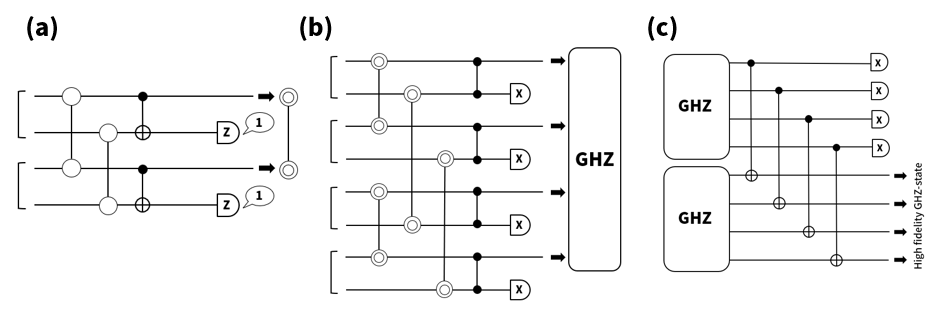}
    \caption{\label{fig15} \raggedright (a) Bell states purification of EPL protocol. (b) GHZ-state distillation. (c) Purification process to get higher fidelity GHZ-states as in \cite{Nickerson2014_s}.} 
\end{figure}

Generation of distributed surface code in a 2D regular quadrilateral grid of memory cells as discussed in the original proposal \cite{Nickerson2014_s} is schematically shown in Fig.~\ref{fig15}. The first stage of that protocol consists in distribution and purification of the Bell states between neighbouring nodes (Fig.~\ref{fig15}(a)). Purification was proposed to be performed according to the extreme photon loss (EPL) protocol \cite{Campbell2008_s}. Then the GHZ states are distilled as shown in Fig.~\ref{fig15}(b), and purified, Fig.~\ref{fig15}(c).

In the main text, we show that our procedure of GHZ state generation can significantly simplify the original proposal by reducing the entire protocol to two stages: direct generation of GHZ states when appropriate set of detectors records photodetection events, and purification. We can propose to perform the latter by generalizing the EPL protocol, shown in Fig.~\ref{fig15}(a), as discussed in the main text.   

Here we discuss the details of the procedure of GHZ state purification from the mixed state
\begin{equation}
    \label{actual_ss}
    \rho = \alpha\ketbra{\Phi} + \sum \beta_i\ketbra{\phi_i},
\end{equation}
where $\alpha+\sum \beta_i=1$ and \{$\ket{\phi_i}$\} are undesirable states. We notice that different configurations of detectors signalize mixed GHZ states up to different local unitary transformations. The local transformations can correct the first component of the equation (\ref{actual_ss}), but the undesirable contributions are not transformed to the same set. Therefore the purification procedure requires observing a given patterns of detection events which reduces the rate by a constant factor.  

\begin{figure}[h]
    \centering
    \includegraphics[width=0.6\textwidth]{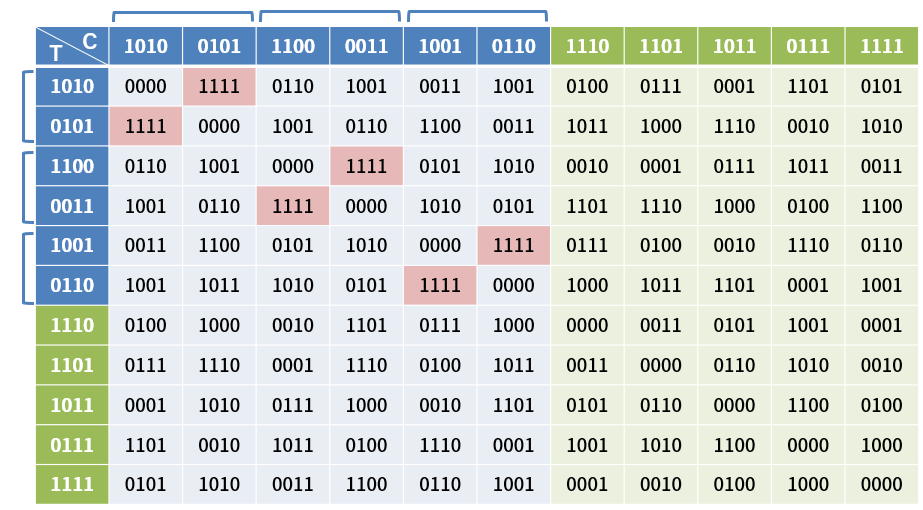}
    \caption{\label{fig17} \raggedright The first row and column of the table show the terms from two copies of the mixed state (\ref{actual_ss}), respectively. Here the first copy serves as the control system (C), while the second as the target (T). The inner part of the table shows values of the target states' terms after a set of CNOT. 
    The measurement $1111$ is the only result which heralds the elimination of unwanted terms denoted in the last columns and rows 
 (in green), and hence, the success of the GHZ state purification. We distinguished pairs of states that correspond to the same set of detectors generating the mixed states in our protocol.} 
\end{figure}

The one step of the purification protocol requires two copies of $\rho$ -- see Fig.~5 of the main text. One is treated as the source of control qubits, the second as the target. The corresponding terms of the control and target qubits are shown in the first row and column, respectively, of the table in Fig.~\ref{fig17}. The inner part of the table shows the values of the target qubits after CNOT gates are applied. These qubits are measured. Note that the only measurement result that eliminates all undesired terms is $1111$. Therefore this measurement leads us to the purified version of GHZ state.  This procedure is a generalization of the extreme loss protocol (EPL) shown in Fig.~\ref{fig15}(a).



\end{document}